\documentclass[aps,pra,twocolumn,superscriptaddress,showkeys]{revtex4}
\usepackage{graphicx}
\usepackage{dcolumn}
\usepackage{color}
\usepackage{bm}
\usepackage{amssymb}
\usepackage{amsmath}
\begin{document}
\title{Resonance enhancement of two photon absorption by magnetically trapped atoms in strong rf-fields}

\author{A. Chakraborty}
\email[E-mail: ]{carijit@rrcat.gov.in}
\author{S. R. Mishra}
\affiliation{Raja Ramanna Centre for Advanced Technology, Indore-452013, India.}
\affiliation{Homi Bhabha National Institute, Mumbai-400094, India.}

\begin{abstract}
Applying a many mode Floquet formalism for magnetically trapped atoms interacting with a polychromatic rf-field, we predict a large two photon transition probability in the atomic system of cold $^{87}Rb$ atoms. The physical origin of this enormous increase in the two photon transition probability is due to the formation of avoided crossings between eigen-energy levels originating from different Floquet sub-manifolds and redistribution of population in the resonant intermediate levels to give rise to the resonance enhancement effect. Other exquisite features of the studied atom-field composite system include the splitting of the generated avoided crossings at the strong field strength limit and a periodic variation of the single and two photon transition probabilities with the mode separation frequency of the polychromatic rf-field. This work can find applications to characterize properties of cold atom clouds in the magnetic traps using rf-spectroscopy techniques. 
\end{abstract}

\maketitle

\section{Introduction}

In the recent years, periodically driven quantum systems have been of much interest due to their unique and fundamental features which find applications in a variety of topics in physical sciences. In order to understand these driven systems better, a new theoretical approach known as `Floquet engineering' \cite{Holthaus:2016}, has evolved. This approach provides an opportunity to manipulate the interactions of an atomic system with a periodic driving force while transforming the time dependent interaction Hamiltonian to a `stroboscopic' Hamiltonian \cite{Bukov:2015}. Using Floquet engineering technique, different regimes of atom-field interaction can be studied by setting the field parameters appropriately for the chosen regime.

In this work we have applied Floquet theory on an atomic system trapped in a static magnetic trap and interacting with a strong polychromatic rf-field. Magnetically trapped atoms interacting with an rf-field is an extensively studied system due to its application in widely known cooling mechanism `evaporative cooling' which facilitated the generation of the quantum degenerate state of matter known as Bose Einstein Condensate \cite{Anderson:1995,Yamashita:1999}. Another interesting application of this atomic system exists in trapping of atoms in non-trivial geometries using the rf-dressed potentials arising due to the creation of dressed states resulting from the interaction of atoms with the strong rf-fields \cite{Zobay:2001,Colombe:2004,White:2006,Lesanovsky:2007}. All these cooling and trapping schemes based on the use of rf-radiation have mostly used the monochromatic rf-radiation \cite{Heathcote:2008,Morizot:2006, Chakraborty:2016}, where the effect of a polychromatic rf-field is considered only to a limited extent \cite{Courteille:2006,Hofferberth:2007}. One major hindrance while dealing with the polychromatic rf-field is the breakdown of the widely used rotating wave approximation (RWA). To overcome this limitation, use of the many mode Floquet theory (MMFT) seems an appropriate choice \cite{Ho:1983,Ho:1984,Ho:31:1985,Ho:32:1985,Chu:2004}. 

A theoretical approach based on the MMFT is applied in this article to understand the atom-field interactions between cold $^{87}Rb$ atoms trapped in a quadrupole trap in hyperfine state $|F=2\rangle$ and exposed to a polychromatic rf-field. Due to the presence of the spatially varying static field, the energy separation between different Zeeman sub-levels change, leading to a position dependent absorption of the rf-photons from the resonant component of the applied rf-field. These absorption of the rf-photons induce transitions between different Zeeman sub-levels and depending on the energy gap between the initial and the final states, these transitions can involve either absorption of a single rf-photon (transition between two adjacent Zeeman sub-levels) or a simultaneous absorption of two rf-photons (transition between two Zeeman sub-levels having another intermediate sub-level). More higher order transitions can also be studied by choosing an atomic species with a hyperfine ground state higher than $|F=2\rangle$.

In the results, the transition probabilities between the Zeeman sub-levels has been calculated and a dependence of both the single and two photon transition probabilities on the applied field strength is obtained. The calculated two photon transition probability (peak value obtained $\sim 0.2$) is larger than its typical values in the optical domain (peak value shown $\sim 5\times 10^{-3}$ in \cite{Ho:1983}). The overall increase in the single and two photon transition probabilities has been explained by inspecting the Floquet spectrum for different regimes of the rf-field strength and the analytical explanation agrees well with the numerical results. Further, we note that the prime reason for the large two photon transition probability in our case could be due to the presence of the resonant intermediate state to which the population is excited from the initial ground state, resulting in resonance enhancement effect. The dependence of both, the two photon transition probability and population in the intermediate state, has been studied as a function of the coupling strength and the relevant results are discussed. Among others, one possible application of these large transition probabilities can be the determination of trap frequencies, and hence the trapped cloud temperature, using the rf-spectroscopy techniques \cite{Easwaran:2010}.

\section{Theory}
In this section, we will discuss only the key components of the many mode Floquet formalism, a detail description can be found elsewhere \cite{Ho:1983,Ho:1984}. We define the time-independent atomic Hamiltonian as $H_0=\mathcal{H}-\boldsymbol{\mu}.\textbf{B}_s(\textbf{r})$, where $\mathcal{H}$ is the unperturbed Hamiltonian and $-\boldsymbol{\mu}.\textbf{B}_s(\textbf{r})$ is the static field trapping energy. Here $\boldsymbol{\mu}$ is the magnetic dipole moment and $\textbf{B}_s(\textbf{r})$ is the static magnetic field of form $B_q[x,y,-2z]^{T}$, with $B_q$ being the radial field gradient. Here, the Larmor precision frequency $\omega_L$ associated with this trap is position dependent and can be written as,
\begin{equation}\label{eq:omega_L}
\omega_L(\textbf{r})=\frac{g_F\mu_BB_q}{\hbar}\sqrt{x^2+y^2+4z^2}.
\end{equation}
The multi-mode rf-field can be described as $\textbf{B}(t)$ consisting the monochromatic field modes $\textbf{B}_k$ with frequencies $\omega_k=\omega_0+k\omega_r$, with $\omega_0$ and $\omega_r$ being the central frequency and the mode separation frequency respectively. Assuming that the modes lying higher than a certain value $k>N$ are of negligible strength, the mode index $k$ can be written as $k=0,\pm 1,\pm 2,...,\pm N$, resulting in $2N+1$ modes in the spectrum. Hence, the Schr\"{o}dinger equation which govern the dynamics of this system can be written as, 
\begin{equation}\label{eq:schrodinger}
i\hbar\frac{\partial\psi}{\partial t}=H\psi
\end{equation}
with 
\begin{equation}
H=H_0-\boldsymbol{\mu}.\textbf{B}(t)=H_0-\sum_{k=-N}^{k=N}\boldsymbol{\mu}.\textbf{B}_k\Re[e^{i\omega_kt}].
\end{equation}
To yield a proper representation of the Hamiltonian $H$, we employ a tensor basis set comprising the unperturbed atomic basis set $\{|\alpha\rangle\}$ and two independent Fourier basis sets $\{|n\rangle\}$ and $\{|m\rangle\}$ for the frequencies $\omega_0$ and $\omega_r$ respectively. The composite basis set can be defined as,
\begin{equation}
\{|\alpha nm\rangle\}=\{|\alpha\rangle\}\otimes\{|n\rangle\}\otimes\{|m\rangle\}.
\end{equation}
By representing the Hamiltonian $H$ in this basis set, we obtain the Floquet Hamiltonian $H^F$ and the time dependent Eq. (\ref{eq:schrodinger}) gets transformed to a `stroboscopic' eigen value equation as,
\begin{equation}\label{eq:HF}
\sum_{\beta}\sum_{n'}\sum_{m'}\langle\alpha nm|H^F|\beta n'm'\rangle\langle\beta n'm'|\epsilon\rangle=\epsilon\langle\alpha nm|\epsilon\rangle,
\end{equation}
where $\epsilon$ is the eigen-energy and $|\epsilon\rangle$ is the corresponding eigen-vector. The matrix elements of the Floquet matrix $H^F$ can be determined as,
\begin{multline}
\langle\alpha nm|H^F|\beta n'm'\rangle=\epsilon_\alpha\delta_{\alpha,\beta}\delta_{n,n'}\delta_{m,m'}+\\ \sum_{k=-N}^{N}\Omega_{\alpha\beta}^{(k)}\left(\delta_{n+1,n'}\delta_{m+k,m'}+\delta_{n-1,n'}\delta_{m-k,m'}\right)+\\(n\omega_0+m\omega_r)\delta_{\alpha,\beta}\delta_{n,n'}\delta_{m,m'},
\end{multline}
where $\delta_{i,j}$ is the Kronecker delta function. The unperturbed eigen-energies $\epsilon_{\alpha}$ and the different coupling strengths $\Omega_{\alpha\beta}^{(k)}$ between the different atomic states are described as,
\begin{subequations}
\begin{align}
\epsilon_\alpha&=\langle\alpha|H_0|\alpha\rangle,\\
\Omega_{\alpha\beta}^{(k)}&=-\frac{1}{2}\langle\alpha|\boldsymbol{\mu}.\textbf{B}_k|\beta\rangle.
\end{align}
\end{subequations}

The time evolution of the composite atom-field system of Eq. (\ref{eq:schrodinger}) can be described by the matrix form of the time evolution operator $U_{\beta\alpha}(t,t_0)$ described in terms of the Floquet Hamiltonian as,
\begin{multline}\label{eq:U}
U_{\beta\alpha}(t,t_0)=\sum_{\{n,m\}=-\infty}^{\{n,m\}=\infty}\langle\beta nm|\exp[-iH^F(t-t_0)]|\alpha 00\rangle\\\times\exp[i(n\omega_0+m\omega_r)t].
\end{multline}
The transition probability between unperturbed atomic states can be calculated using the time evolution operator of Eq. (\ref{eq:U}) by,
\begin{equation}
P_{\alpha\rightarrow\beta}(t,t_0)=|U_{\beta\alpha}(t,t_0)|^2.
\end{equation}
An average over the initial time $t_0$, with a fixed elapsed time ($t-t_0$), leads to the transition probability, 
\begin{multline}
\begin{aligned}
P_{\alpha\rightarrow\beta}(t-t_0)
&=\sum_{nm}|\langle\beta nm|\exp[-iH^F(t-t_0)]|\alpha 00\rangle|^2\\
&=\sum_{nm}\sum_{k_1,k_2}\sum_{\epsilon n'm'}|\langle\beta k_1k_2|\epsilon n'm'\rangle|^2\\
&\hspace{1cm}\times\langle\alpha nm|\epsilon n'm'\rangle\langle\epsilon n'm'|\alpha 00\rangle\\
&\hspace{1cm}\times\exp[i(n\omega_0+m\omega_r)t_0].
\end{aligned}
\end{multline}
A further average over the elapsed time $(t-t_0)$ provides the time-averaged transition probabilities between two atomic states $|\alpha\rangle$ and $|\beta\rangle$ in terms of only the eigen-states of the Floquet Hamiltonian $H^F$ as \citep{Ho:31:1985},
\begin{equation}\label{eq:pab}
\bar{P}_{\alpha\rightarrow\beta}=\sum_{n,m}\sum_{\epsilon n'm'}|\langle\beta nm|\epsilon n'm'\rangle\langle\epsilon n'm'|\alpha 00\rangle|^2.
\end{equation}

In standard perturbation theory approach, the single and two photon transition probabilities are calculated using first and second order perturbation theory \cite{Grynberg:book}. In contrast, the transition probability obtained from MMFT in Eq. (\ref{eq:pab}) can be utilized to determine single and multi-photon transition probabilities without a prior specification of the total number of photons involved in the transition process. The obtained transition probability $\bar{P}_{\alpha\rightarrow\beta}$ is equivalent to the one obtained using an infinite order perturbation theory \cite{Chu:1977,Ho:1983}. 

As the energy levels under consideration are within the Zeeman sub-levels of the ground hyperfine state $|F=2\rangle$, transitions between these levels are electric dipole forbidden, and the spontaneous decay rate is negligibly small. Hence, the steady state density matrices $\bar{\rho}_{\alpha\beta}$ can be obtained following the approach described in \cite{Ho:1986} without considering the relaxation processes involving spontaneous emission. Following that approach, expression for the long time averaged populations can be written as,
\begin{equation}\label{eq:rho}
\bar{\rho}_{\alpha\alpha}=\sum_{n,m}\sum_{\epsilon n'm'}|\langle\alpha nm|\epsilon n'm'\rangle\langle\epsilon n'm'|000\rangle|^2.
\end{equation}

The single and two photon transition probabilities along with the population of individual energy levels shown in the next section are determined using the expressions given in Eq. (\ref{eq:pab}) and Eq. (\ref{eq:rho}), respectively, with the values of the parameters chosen as close as possible to their typical experimental values.

\section{Results and Discussions}
In the numerical simulations, we have assumed a Gaussian distribution of field strengths for various modes as a function of the mode frequency. In the calculations, using the Floquet formalism described above, we tune the coupling strength of the central mode $\Omega_0$ and the separation frequency between consecutive modes $\omega_r$ to access different regimes of the interaction. The atomic system considered is comprising of $^{87}Rb$ atoms trapped in a quadrupole magnetic trap in the hyperfine state $|F=2\rangle$ with a radial field gradient $B_q=100\ G\ cm^{-1}$. The polychromatic rf-field used in this study contains $8$ side modes and one central mode with frequency $\omega_0=2\pi\times 2$ MHz. The values of the parameters such as $\omega_0$, $\omega_r$, $B_q$ and $\Omega_0$ are chosen such that experiments can be performed conveniently. 

In order to gain an insight into the various physical processes, the eigen-energy spectrum (or ``Floquet spectrum") is obtained by diagonalizing the Floquet Hamiltonian $H^F$. The eigen-energies in Floquet theory are called `quasi-energies', akin to the `quasi-momentum' for the momentum of free electrons in the Bloch theory of lattices. In the Floquet formalism, the eigen-energies can be written as, 
\begin{equation}\label{eq:totalenergy}
\epsilon_{\alpha nm}=\epsilon_{\alpha 00}+n\hbar\omega_0+m\hbar\omega_r,
\end{equation}
where $n$ and $m$ are integers representing different orders of extension of the Fourier basis states for $\omega_0$ and $\omega_r$ respectively, denoting the energy band indices. Here, $\epsilon_{\alpha 00}$ represents the eigen-energy of the central Floquet spectrum for the indices $n=0$ and $m=0$.

As the index $n$ belongs to the set of integers, the eigen-energy values $\epsilon_{\alpha nm}$ increase indefinitely with $n$, while preserving the fundamental energy structure associated with $\epsilon_{\alpha 0m}$. Therefore, similar to the Bloch theory, we utilize this repetition of energy structure to identify the Floquet-Brillouin zone (FBZ) as the fundamental manifold of eigen-energies corresponding to $n=0$. Hence, our discussion of the results are limited upto the boundary of the first FBZ defined by the eigen-energy $\epsilon_{\alpha nm}$ lying in the energy interval $\left[-\hbar\omega_0/2, \hbar\omega_0/2\right]$. The index $m$ relates to the polychromaticity of the radiation field and thus restricted to $(2N+1)$ values, mirroring the total number of modes present in the spectrum. Hence, inside the FBZ, a total of $(2N+1)$ individual eigen-energy sub-manifolds will emerge corresponding to different $m$ values.

\begin{figure}[t]
\includegraphics[width=8.5cm]{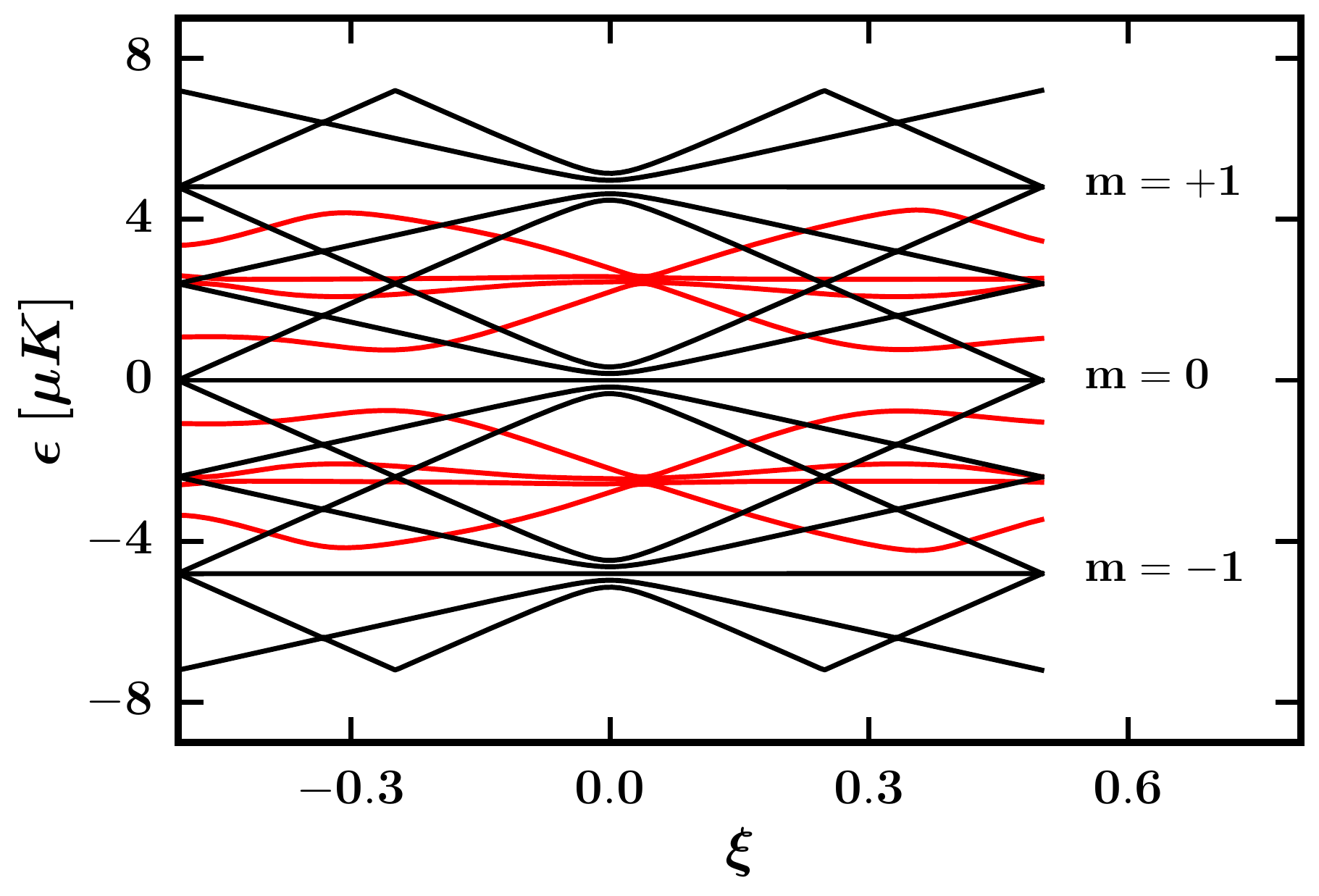}
\caption{\label{fig:avcro} (Color online) Eigen-energy levels as a function of the dimensionless scale $\xi$ for three different sub-manifolds (\textit{i.e.} $m=0,\pm 1$) of the first FBZ (\textit{i.e.} $n=0$). The black curves correspond to the eigen-energies for $\Omega_0=2\pi\times 2$ kHz and the red curves correspond to $\Omega_0=2\pi\times 400$ kHz. At $\xi=0$, the black curves form primary avoided crossing and the red curves form secondary avoided crossing.}
\end{figure}

The eigen-energies inside the FBZ are uniquely defined and cannot cross each other as a result of the Von-Neumann-Wigner non-crossing rule. According to this rule, the crossing of eigen-energies are forbidden while only one single parameter is varied in the whole parameter space in which the Hamiltonian is spanned. Therefore, a spatial variation in the energy will essentially result in an avoided crossing at all the points where any of the field modes become resonant with the spatially varying Larmor precision frequency $\omega_L(\textbf{r})$. In Fig. \ref{fig:avcro}, the variation of eigen-energies in the first FBZ for three sub-manifolds ($m=0,\pm 1$) has been plotted as a function of $\xi(\textbf{r})$ to show the formation of the avoided crossings at very low (\textit{i.e.} $\Omega_0=2\pi\times\ 2$ kHz shown by black curves) and high (\textit{i.e.} $\Omega_0=2\pi\times\ 400$ kHz shown by red curves) values of the coupling strength. This choice of $\xi(\textbf{r})$, a dimensionless parameter defined as $\xi(\textbf{r})= (\omega_L(\textbf{r})-\omega_0)/\omega_r$, as a spatial parameter is to take into account the variation in the relevant frequency components involved in the process which makes the interpretation of the results easier. The coupling strengths are chosen to represent two separate regimes in which the formed avoided crossings are different in nature and provides the best overall visibility of the Floquet eigen-spectrum. 

In case of a very small coupling strength (\textit{e.g.} $\Omega_0\approx 2\pi\times\ 2$ kHz), the avoided crossings formed at the center (\textit{i.e.} $\xi(\textbf{r})=0$) belong to a particular sub-manifold with a single $m$ value as shown by the black curves in Fig. \ref{fig:avcro}. The separation between the energy levels in these avoided crossings depend on the coupling strengths as a consequence of the conservation of energy. Hence, at the intermediate coupling strengths, an increase in $\Omega_0$ increases the eigen-energy values monotonically. However, at a larger coupling strength (\textit{e.g.} $\Omega_0\approx 2\pi\times\ 400$ kHz) a situation can arise where the energy levels from adjacent sub-manifolds (with different m values) can come close to each other to form a different kind of avoided crossing known as the `secondary avoided crossing'. The secondary avoided crossings formed by levels with different $m$ values are shown by the red lines in the Fig. \ref{fig:avcro}. The energy levels in the secondary avoided crossing can be arbitrary close to one another due to the fact that the total energy is already conserved by the primary avoided crossings.

\begin{figure}[b]
\includegraphics[width=8.5cm]{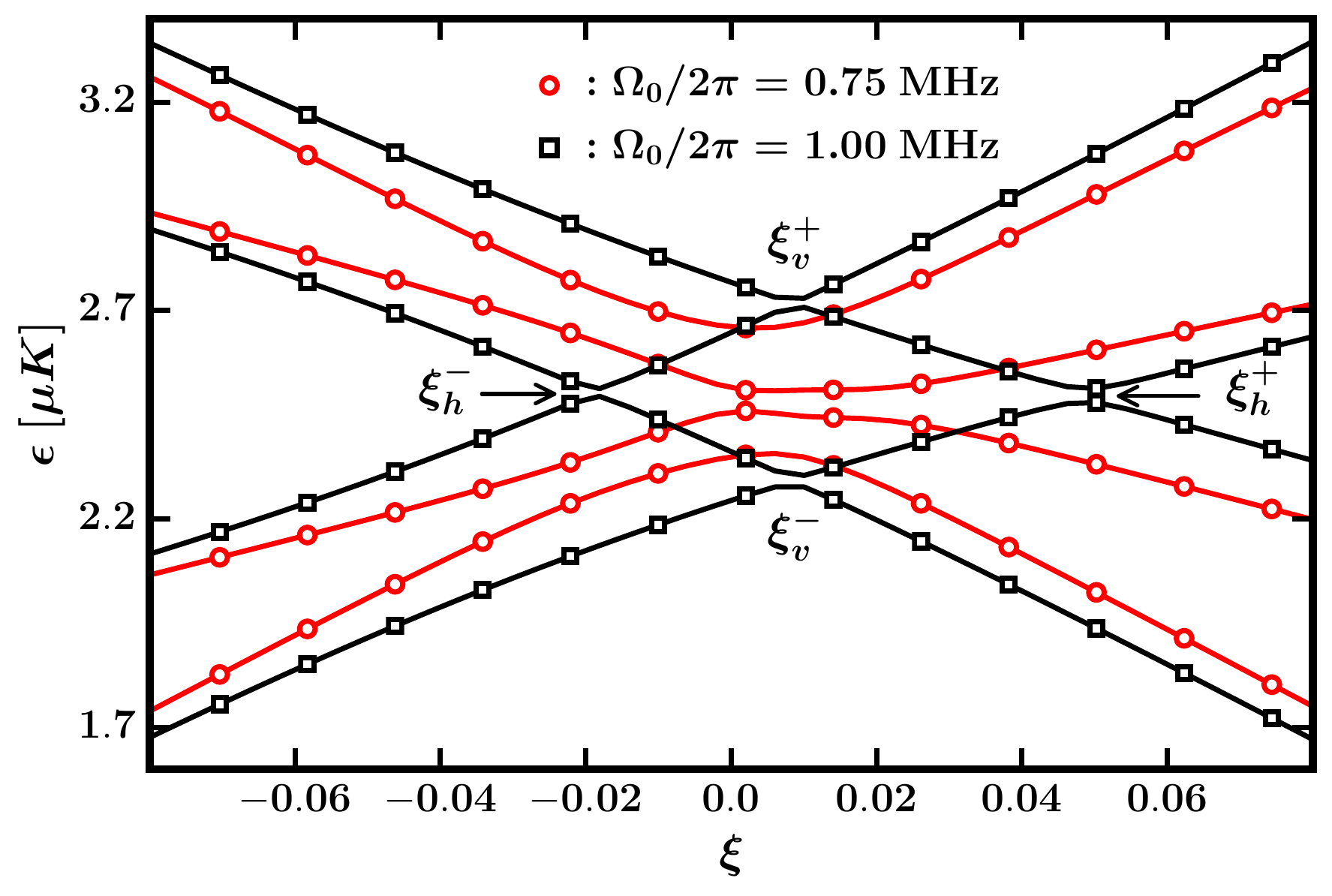}
\caption{\label{fig:Brf_split} (Color online) Variation of the eigen-energy levels as a function of the scaled spatial variable $\xi$ for coupling strength $\Omega_0=2\pi\times\ 0.75$ MHz (red curve) and $\Omega_0=2\pi\times\ 1.00$ MHz (black curve). The splitting of the single avoided crossing into two horizontal avoided crossings $\xi_h^+$, $\xi_h^-$ and two vertical avoided crossings $\xi_v^+$, $\xi_v^-$ is shown in the plot.}
\end{figure}

The secondary avoided crossings were studied by increasing the coupling strength further. In Fig. \ref{fig:Brf_split} the energy levels associated with a secondary avoided crossing are shown with the red curve for a coupling strength of $\Omega_0=2\pi\times 0.75$ MHz. At a higher coupling strength $\Omega_0=2\pi\times 1$ MHz, the single secondary avoided crossing split into four different avoided crossings which unfurl in different directions as shown by the black curves in Fig. \ref{fig:Brf_split}. The vertically formed avoided crossings at $\xi_v^+$ and $\xi_v^-$ retain their spatial position while getting separated in the energy values. The horizontally formed avoided crossings shift in two opposite spatial positions at $\xi_h^-$ and $\xi_h^+$. This horizontal spread of these avoided crossings is asymmetric with respect to the initial crossing position at $\xi(\textbf{r}=0)$. This asymmetric spread of the split avoided crossings is due to the inherent asymmetry of the eigen-values of the static-field Hamiltonian $H_0$ in the quadrupole trap. 

\begin{figure}[b]
\includegraphics[width=8.5cm]{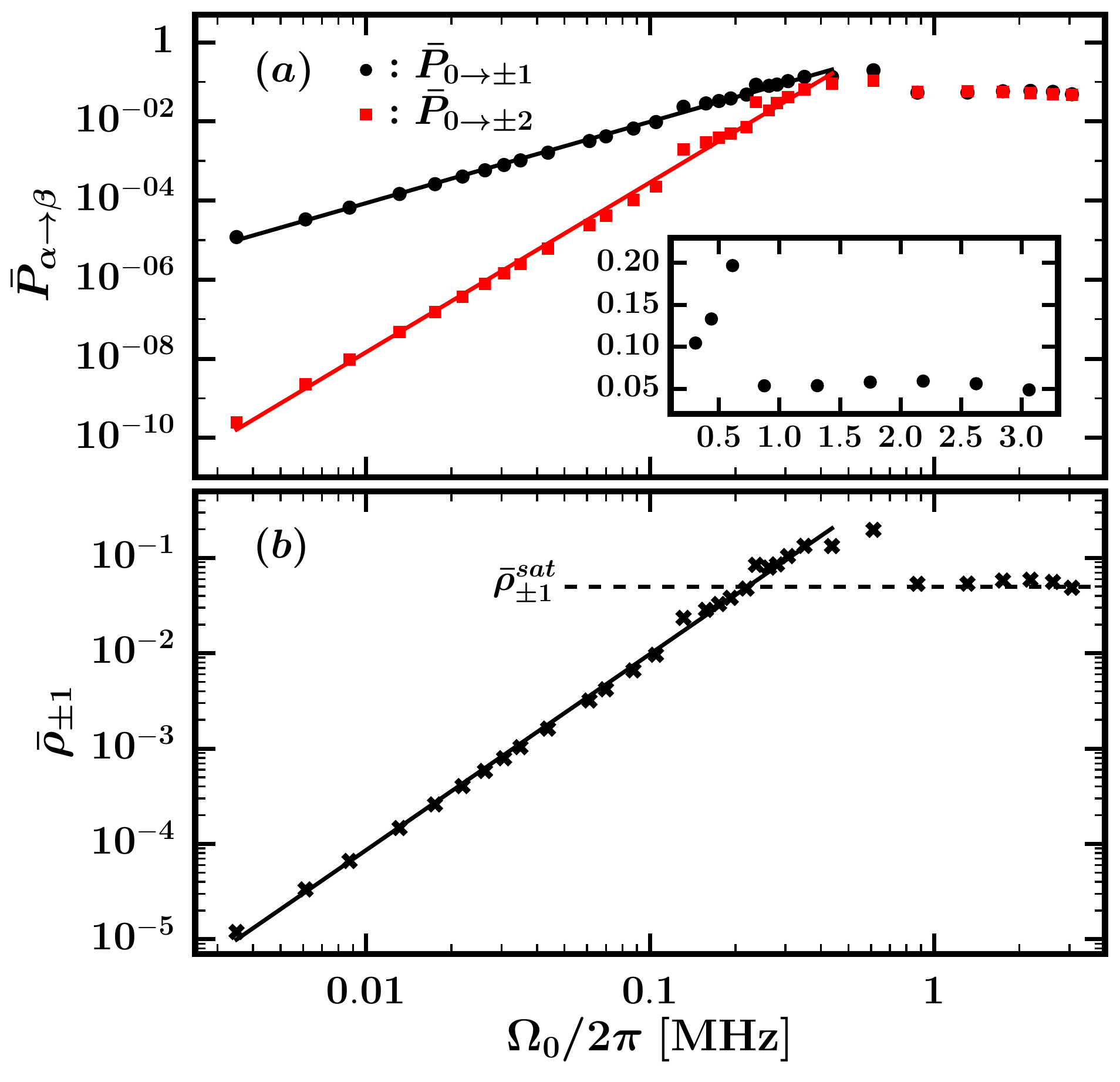}
\caption{\label{fig:Brf_P12} (Color online) (a): Single (black) and two (red) photon transition probabilities and (b): intermediate state populations, as a function of the coupling strength $\Omega_0/2\pi$. The solid lines are linear fit to the numerical values. The inset in (a) shows the magnified view of the variation of the single photon transition probability with the coupling strength in a linear scale. The dashed line in (b) marks the saturation value of intermediate states populations, \textit{i.e.} $\bar{\rho}_{\pm 1}^{sat}$.}
\end{figure}

To study the effect of the transformation of primary avoided crossings to secondary avoided crossings on the transition probabilities between different atomic states, we varied the peak coupling strength $\Omega_0$ in a wide but experimentally feasible regime and obtained the single and two photon transition probabilities and the results are shown in Fig. \ref{fig:Brf_P12} (a). Initially, with a smaller coupling strength ($\Omega_0\ =2\pi\times\ 2$ kHz), the two photon transition probabilities are 5 orders of magnitude smaller than that of the single photon transition probabilities ($\approx 10^{-5}$). The avoided crossings formed at this coupling strength are of primary type as shown by the black lines in Fig. \ref{fig:avcro}. The large separation between the energy levels results in the smaller values of transition probabilities between the atomic states in this regime. As the coupling strength is increased, both the transition probabilities initially grow linearly but with different rates. Finally at the $\Omega_0\ =2\pi\times\ 0.75$ MHz, both the transition probabilities become large and almost comparable to each other. This enhancement in transition probabilities is due to the fact, that all the avoided crossings are completely transformed to the secondary type of avoided crossing with very narrow energy separation between the energy levels, as shown by the red curves in Fig. \ref{fig:avcro}. As the transition probabilities in the studied atom-field composite system peak due to the very small separation between the energy levels, these transitions can be considered as Majorana like transition with finite energy offset and non-zero level separation energy. The sharp decrease in the transition probabilities beyond $\Omega_0= 2\pi\times 0.75$ MHz, as shown in the inset of Fig. \ref{fig:Brf_P12} (a), is due to the splitting of the avoided crossing which is shown in Fig. \ref{fig:Brf_split}. This splitting of the secondary avoided crossing affects the resonance condition to reduce the overall transition probability. 

In order to study the effect of variation of the coupling strength on the intermediate states population, $\bar{\rho}_{\pm 1}$ has been evaluated using Eq. (\ref{eq:rho}) and the results are shown in Fig. \ref{fig:Brf_P12} (b). As evident from the figure, the population $\bar{\rho}_{\pm 1}$ increases linearly below $\Omega_0=2\pi\times 0.75$ MHz, and then leaps down to a nearly constant value $\bar{\rho}_{\pm 1}^{sat}$. The similar dependence of $\bar{\rho}_{\pm 1}$ and $\bar{P}_{0\rightarrow\pm2}$ on the coupling strength shows the direct correlation between the single photon resonance and the two photon transition probability. The saturation of the populations $\bar{\rho}_{\pm 1}$ beyond $\Omega_0=2\pi\times 0.75$ MHz also leads to the saturation of the two photon transition probability. These results establish the effect of resonance enhancement on two photon transition probabilities. 

We note here that the two photon transition probabilities obtained here (peak value $\sim 0.2$) are much larger than their counterpart in the optical domain (peak value $\sim 5\times 10^{-3}$ \cite{Ho:1983}). One of the key difference between the interaction of atoms in rf and optical  frequency domain is the associated energy scales relevant to a specific transition. In case of our composite atom-field system, the coupling strength (orders of MHz) is comparable with respect to the energy separation (orders of MHz) between the Zeeman sublevels. This is in contrast with the optical domain transitions where the energy level separation (orders of THz) is much higher than the typically applied coupling strengths (orders of MHz).

\begin{figure}[t]
\includegraphics[width=8.5 cm]{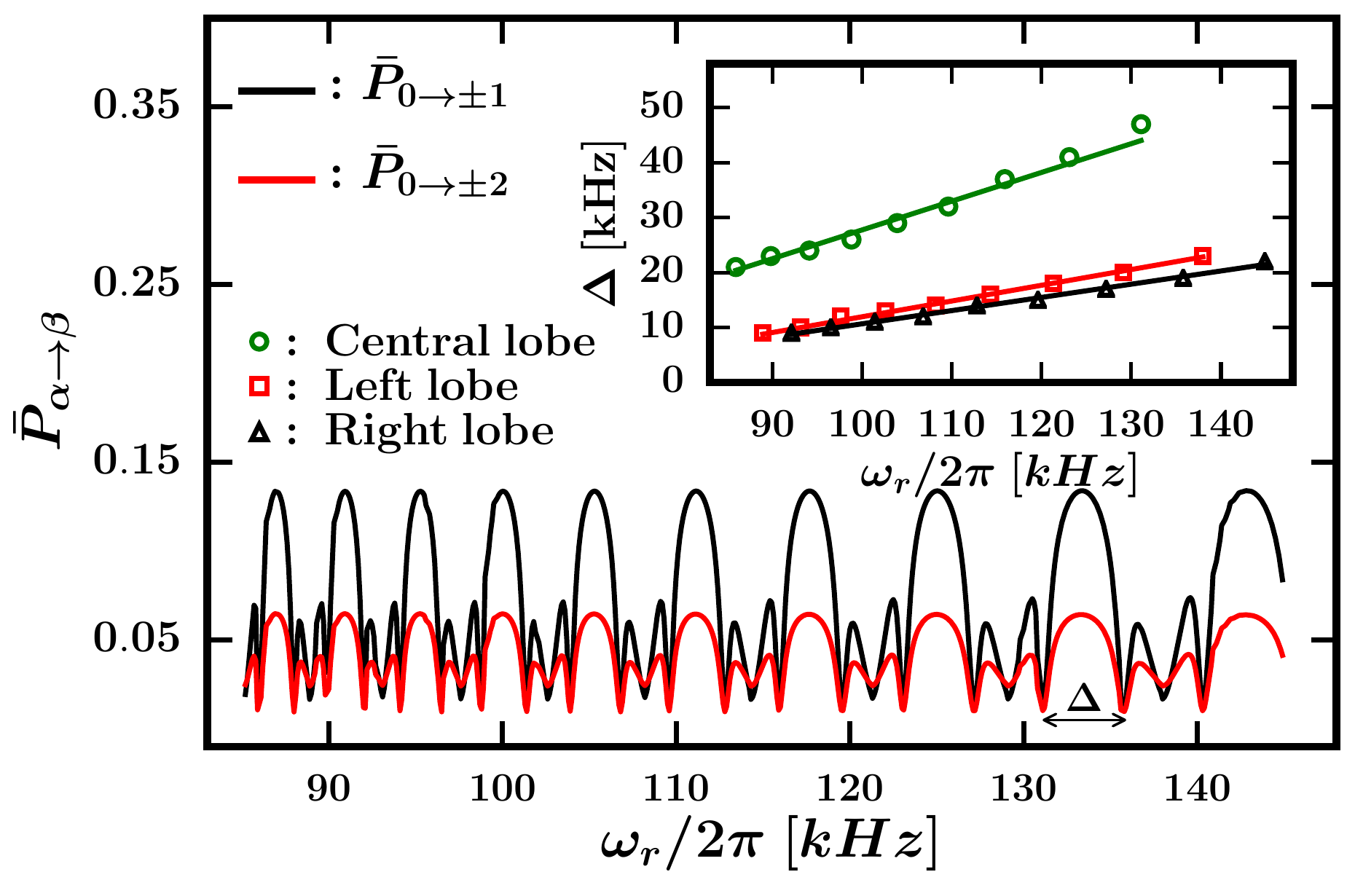}
\caption{\label{fig:wr} (Color online) Variation of the single (black curve) and two (red curve) photon transition probabilities as a function of the separation frequency $\omega_r$ at the central resonance $\xi=0$. The inset shows the width ($\Delta$) of the single photon transition probability lobes as a function of the separation frequency $\omega_r$. In the inset, circles, squares and triangles denote the width of the central, left and right lobes in the transition probability. The solid lines depict the linear fits.}
\end{figure}

Further, we have also studied the effect of variation in $\omega_r$ on the transition probabilities. The results of this variation are shown in Fig. \ref{fig:wr}. Both the transition probabilities periodically repeat a basic pattern which consists one central lobe accompanied by two smaller side lobes. The width of these transition probability lobes ($\Delta$) increases with the separation frequency $\omega_r$, whereas the peak height of the lobes remain same. The width of the individual lobes are plotted in the inset of Fig. \ref{fig:wr} where a linear increase in the width values is evident. The width of the smaller lobes are not equal with each other and the difference also increases linearly as a function of $\omega_r$ as shown by the curves with squares (left lobe) and triangles (right lobe). A closer inspection of the Floquet spectrum enables us to identify the source of the formation of these transition probability lobes. It is already known, that at the position of the avoided crossings, the transition probability drops significantly as compared to that at the positions of the real crossings. Here, the variation in $\omega_r$ leads to a periodic variation in the avoided crossing positions, which finally results in the periodic variation in the transition probabilities. Hence, the recurring lobes of the transition probability are only an outcome of the periodic spatial shifting of the avoided and real crossings as a function of the separation frequency $\omega_r$. This specific variation of the transition probability on the mode separation frequency $\omega_r$ can be exploited to either suppress or excite transitions between different states by only tuning the $\omega_r$ to a desired frequency value.

\section{Conclusion}
Using many mode Floquet theory, we have calculated the eigen-energy spectrum, transition probabilities and populations of atomic states for cold $^{87}Rb$ atoms trapped in a quadrupole trap while being exposed to a polychromatic rf-field. The transformation of the primary avoided crossings into the secondary avoided crossings has been observed by varying the coupling strengths, which in turn resulted in redistribution of populations in the atomic states along with large transition probabilities. The observed transition probabilities have also shown periodic variation as a function of the mode separation frequency and can be employed to either suppress or excite specific transitions between different states. Possible applications of this study can be in determination of the trap frequencies and atom cloud temperatures using rf-spectroscopy techniques as well as in studies of higher order non-linear processes like Anderson localization.

\section*{Acknowledgements}
We acknowledge Charu Mishra for a careful reading of the manuscript. A. Chakraborty acknowledges the financial support by Raja Ramanna Centre for Advanced Technology, Indore under the HBNI programme.

%\bibliographystyle{apsrev}
%\bibliography{Reference}

\end{document}